\documentclass[twocolumn,superscriptaddress,prl]{revtex4}
\usepackage{amssymb}
\usepackage{graphicx}
\usepackage{dcolumn}
\usepackage{bm}
\usepackage{amsmath}

\usepackage{wasysym}
\usepackage[normalem]{ulem}
\usepackage[usenames,dvipsnames]{color}
\begin{document}

\title{Catalytic Buffering for Optimal Scheduling of Self-Replication}

\author{R. Pugatch}
\affiliation{Simons Center for Systems Biology, School of Natural Sciences, Institute for Advanced Study, Princeton, NJ 08540, USA}
\email{rami.pugatch@gmail.com}

\begin{abstract}
We study the scheduling problem of a self-replicating factory. We show that by maintaining a sufficiently large inventory of intermediate metabolites and catalysts required for self-replication, optimal replication times can be achieved by a family of random scheduling algorithms that are biochemically feasible, for which catalysts never idle if they can perform de-novo bio-synthesis.
Optimally scheduled self-replication is facilitated by allowing several production lines to run in parallel. The excess inventory of catalysts and substrates decouples these lines, while dynamical balancing tunes average and variance completion, resulting in an overall universal distribution for the replication times belonging to the generalized extreme value family. We discuss biological implications and postulate that bacteria that are tuned by evolution for fast replication employ this natural scheduling strategy to achieve optimal asymptotic growth rates by stoichiometrically balancing the amount of work in progress thus globally controlling the number of parallel basic self-replicating units within them. Analysis of recently measured data of \textit{E. coli} growth in rich media shows data-collapse on a single universal curve consistent with our prediction, suggesting wild type \textit{E. coli} optimally schedule its replication.
\end{abstract}

\maketitle

An \textit{E. coli} bacterium is a remarkably efficient self-replicating organism. Given the right conditions, this rod-shaped bacterium will consume external metabolites and grow, adding new membrane-bound volume, while concurrently replicating its content composed chiefly from the translation-transcription machinery, metabolism, and DNA. After all the essential elements have been replicated and spatially segregated, \textit{E. coli} splits by completing the construction of the division plane, approximately midway between the old and the new poles. The two approximately identical copies can both continue to self-replicate, as long as the permissive conditions persist.

The study of self-replication as an industrial process was pioneered by John von Neumann \cite{VN1}. In his first, less known model --- the kinematic self-replicator, he envisioned a room full of elementary parts and a non-trivial self-replicating machine that assembles a copy of itself by consuming these parts as substrates. The main goal of von Neumann was to understand how a physical system can become more complex over time. Motivated by the introduction of the universal Turing machine to the theory of computation, he introduced the concept of a ``universal constructor'' ($U$) --- a machine that can read instructions and translate them into actual assembly of any machine in the factory, including itself, provided all the substrates are available. A self-replicating factory is called non-trivial if it contains such a $U$-machine as one of its components. In contrast, trivial self-replication is a simple autocatalytic process, such as template replication (replication by copying a template) or crystal growth.

In von Neumann's model, replication unfolds as follows \cite{VN2}. First, a new chassis is made, then the universal constructor is triggered to start reading the instructions and assembling all the internal machinery, including itself. However, the new factory is not functional until a copy of the instructions is made. To keep the design simple, von Neumann suggested that the instructions should not instruct their own replication but rather be template-replicated by a dedicated machine, $R$ (produced by $U$), that is triggered upon the completion of the instruction translation phase. Remarkably, these observations were made prior to the discovery of DNA, DNA replication, or the ribosome.

While the emphasis of von-Neuman's model is on the logical design of such a factory, here we study the temporal organization of self-replication. The simplest temporal organization is serial. Production is divided into separate steps, which do not overlap in time. There are two ways to enhance the speed of replication. One is to keep the serial temporal organization and speed-up all the production steps. The second is to parallelize production.

The problem of parallelizing a production process given a set of production tasks, their temporal precedence constraints, and a finite amount of resources and processing units, is known as the scheduling problem.
In a typical scheduling problem, a set of processing units has to perform a certain set of tasks with a prescribed partial temporal ordering, such that the completion time of the last task is minimized.

Here, we study the scheduling problem of a self-replicating bacterial cell. A coarse grained picture of a bacterial cell is presented, and its process graph is introduced. The concept of a project graph, well-known in system engineering is invoked to represent the temporal precedence constraints that exist among all the de-novo synthesis tasks that define the project.
We introduce the project graph of a self-replicating and balanced factory and define the ``replicative buffer'' --- the number of basic self-replicating units within a cell. This leads to the explanation of how a replicative buffer greater or equal to one allows a large class of random scheduling algorithms known as list scheduling algorithms \cite{ListScheduling1} to obtain optimal completion times. We derive the distribution of optimal completion times for a special type of project graph representing balanced production. Finally, we analyze two recent data-set of \textit{E. coli} growing fast in a rich medium \cite{jun,Elf}, and show that the data collapses into our predicted optimal universal curve upon proper scaling, suggesting that \textit{E. coli} in optimal growth conditions is optimally scheduling its replication.

\emph{The cell as an autocatalytic cycle.--} A unique feature of a self-replicating factory is \emph{closure}. In a production line, processing units convert raw material into products while also consuming free energy. In a self-replicating factory, the products are the processing units, thus when all the processing units complete their tasks, each processing unit is present in duplicate (or more). To produce a processing unit may require several other processing units, self included. A second feature is \emph{essentiality} --- each processing unit is required by at least one other reaction that produces a different type of processing unit.

In a cell, prominent examples for processing units include protein catalysts, RNA and DNA polymerases (RNAP, DNAP) and ribosomes, since like processing units in a factory, they are required for a production task  (bio-synthesis), they consume input materials and free energy, they are not consumed during the process yet they are essential for its successful completion.

\begin{figure}
\centering
\includegraphics[width=1 \columnwidth]{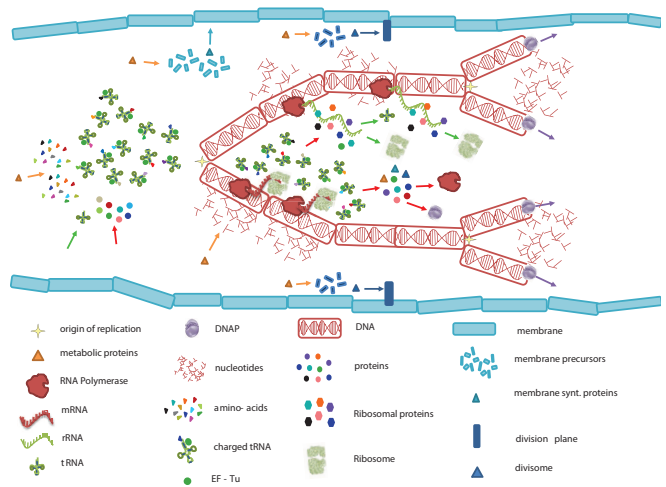}\\
\caption{(Color online) Coarse-grained schematic to demonstrate the closure property of a bacterial cell. Each element in the figure is a representative member of a family of similar molecules sharing the same function. Arrows represent de-novo synthesis. Orange arrows represent metabolic synthesis, with the exception of light and dark blue arrows that represent membrane and division plane synthesis, respectively. Synthesis of tRNA and self-assembly of ribosomes is represented by green arrows. Protein synthesis and self assembly of proteins into protein complexes are represented by red arrows. All proteins are produced by pre-existing ribosomes. All ribosomes are self assembled from rRNA and ribosomal proteins (represented by colored hexagons). All RNA forms are transcribed by RNA Polymerase. While it is possible to further deepen the level of description, any additional functional group not present in this schematic can be produced by one or more groups represented in the schematic. Furthermore, each functional group represented in the figure is essential for the production of members of another group. Note that the synthesis of the membrane is essential to all the internal processes; in the absence of newly formed membrane-bound volume, reactions cannot be sustained due to molecular overcrowding. On the other hand, the complete absence of a membrane (i.e. unlimited volume) might lead to diffusive loss of essential catalysts (in three dimensions).}\label{fig1}
\end{figure}

To illustrate how closure is obtained in a cell, we present a coarse grained schematic of a bacterial cell in Figure (\ref{fig1}). Each symbol in the figure is a representative of a family of functionally related macro-molecules. For example, the DNA replication machinery --- the replisome (the $R$ machine in von Neumann's model),  composed of many types of proteins and protein complexes, is represented in the figure by a single member --- the DNA polymerase (DNAP).

All proteins are synthesized by ribosomes and ribosomes are composed of ribosomal proteins (represented by the colored hexagons in the figure) and ribosomal RNA (rRNA) represented by a green strand. The rRNA is transcribed from DNA by RNA polymerase (RNAP) which is a self-assembled protein complex composed from five different proteins.

To synthesize proteins, ribosomes require a pool of charged transfer RNA (tRNA) and a family of auxiliary proteins such as elongation, initiation and maturation factors, as well as amino-acyl-synthetase proteins that catalyze the tRNA-amino-acids charging. All these auxiliary proteins are represented by the EF-Tu protein (green circle in Figure (\ref{fig1})).
The tRNA is transcribed by RNAP and reach its mature form with the help of some dedicated proteins, (of course,  mRNA is also transcribed by RNAP).  Finally, membrane and division plane synthesis is facilitated by dedicated proteins. All these processes require ribo- and deoxyribo-nucleosides, amino-acids, lipids and oligosaccharide, ATP, GTP and many other metabolites (e.g. $NADH$, and $Mg_2^{++}$) all either synthesized or imported from the outside by the metabolic machinery which is mainly composed of metabolic proteins, represented by orange triangles in Figure (\ref{fig1}). The metabolic proteins catalyze and control the metabolic reactions.

\begin{figure}
\centering
\includegraphics[width=1 \columnwidth]{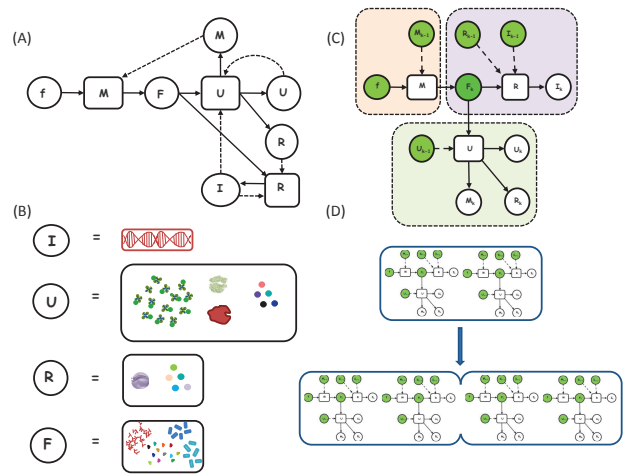}\\
\caption{(Color online) Inset (A) presents the reaction graph described by Eq. (\ref{eq1}). The material components, described by circles with capital letters in the figure, are shown in inset (B) using the same graphical representation as Fig. (\ref{fig1}). For example, the universal constructor $U$ contains ribosomes, RNA polymerases, tRNAs, and auxiliary proteins including initiation, elongation, and release factors as well as a set of amino-acyl-transferase proteins, all required for streamlined production. In inset (C) we illustrate how charging of the catalytic and metabolic pools allow the production lines for $I$, $U$, $M$, $R$ and $F$ to work in parallel. In inset (D) we illustrate a cell with $p=2$ initial basic self-replicating units, that doubles to $p=4$ prior to splitting.}\label{fig2}
\end{figure}

\emph{Introducing the catalytic reaction graph.--} To model the structure of a self-replicating factory, we follow \cite{AutoCatalysis1} and introduce a directed reaction graph $G=\{V,E\}$ called the catalytic reaction graph, with a vertex set $V$ and an edge set $E$. The vertex set $V=M_s \cup R_s$ is comprised of two types of vertices or nodes. A vertex in $M_s$ either represents an input material (substrate) that is required by at least one reaction or an output material produced by at least one reaction. Both types are marked as circles in Fig. (\ref{fig2}) inset (A). A vertex in $R_s$ is a reaction vertex (marked by squares in Fig. (\ref{fig2}), Inset (A)). It represents a reaction that produces new material.

The edge set $E$ contains edges that connect materials into or out from a reaction vertex. The edge set $E=E_R \cup E_c$ is also comprised of two types of edges. An edge in $E_R$ is connecting a consumable input material vertex to the reaction node that consumes it, or connects a reaction node to an output material vertex that is synthesized by that reaction. An edge in the edge set $E_c$ represents a catalyst that is either essential for the reaction to occur.

The material set can be decomposed further to $M_s=F_s \cup C_s$, where $F_s$ is the food set \cite{AutoCatalysis1} --- the set of all consumable material inputs, not necessarily external, and $C_s$ --- the catalytic set, comprised of all the catalysts.

Using this formalism, we can represent a coarse grained model for the cell that is essentially identical to the von-Neuman model. This will facilitate a better characterization of different temporal organizations of self-replication, which we discuss subsequently.

Let $F=F_1 \cup F_2$ be the set of all essential substrates and intermediates that are consumed in at least one de-novo biosynthesis reaction in the cell, and $f$ be the external set of materials (assumed to be fixed and abundant). For example, nucleotides and amino-acids both belong to the set $F$. Let $I$ be the DNA (instruction set). Next, we define $R$ to be the replisome machinery, that catalyzes the template replication of $I$. We define $U$ to be the translation-transcription machinery (the universal constructor) that can read the instruction set and produce a copy of itself as well as all the other processing units.  Finally, we define $M$ to be the set of all processing units that are involved in the biosynthesis process or its control and are not part of $R$ and $U$. For example, $M$ includes among other things all the metabolic proteins, all the transporters, transcription factors, heat shock proteins, and DNA repair proteins.

To complete the definition of the catalytic reaction graph we define the reaction vertex sets. There are three types of reactions: $M_R$ produces the set $F$ from the external materials in $f$. $U_R$ produces all the elements in $M$ and in $U$ from the set $F$, and the reaction $R_R$ produces a copy of $I$ while consuming elements from $F$ (using the existing $I$ as template). The catalytic set $C$ is composed of the subsets $M$, $U$ and $I$.

The following set of reactions summarizes the above description:
\begin{eqnarray}\label{eq1}
   f + M &\underset{k_1}{\longrightarrow}& F + M \\
  \nonumber \alpha F_1 + I_1 + U &\underset{k_2}{\longrightarrow}& I_1 +2U \\
  \nonumber \beta F_1 + I_2 + U &\underset{k_3}{\longrightarrow}& I_2 + U + M \\
  \nonumber \gamma F_1 + I_3 + U &\underset{k_4}{\longrightarrow}& I_3 + U + R \\
  \nonumber F_2 + R + I &\underset{k_5}{\longrightarrow}& 2I +R.
 \end{eqnarray}
where $\alpha+\beta+\gamma=1$ and $\alpha,\beta,\gamma \geq 0$. $\alpha, \beta$ and $\gamma$ are the fractions consumed from  $F_1$ by each of the three reactions: (i) duplicating $U$, (ii) duplicating $M$, and (iii) duplicating $R$. Typically, $\gamma \ll \alpha$ and becuase duplicating the universal constructor involves ribogenesis, $\alpha$ has a dominant metabolic load in rich media i.e. $\alpha > \beta \gg \gamma$.  $F_2$ contains deoxyribo-nucleotides not required by the other reaction.

In Fig. (\ref{fig2}) inset (A) the directed reaction graph of the system described by Eq. (\ref{eq1}) is presented. In inset (B) we describe the content of each material node in the reaction graph (represented by a circle), using the same graphical notation as in Figure (\ref{fig1}). Note that the universal constructor $U$ is composed of the entire translation-transcription machinery (ribosomes. RNA polymerases, tRNAs and all the essential auxiliary proteins).

\emph{From the reaction graph to the project graph.--} The Project Evaluation and Review Technique (PERT) is extensively used in system engineering as a means for scheduling important events in a project, and for estimating the expected project completion time and its distribution \cite{PERT1,PERT2}. A key concept in PERT is the project graph --- a directed acyclic graph whose vertices represent milestone events in the project, marking the initiation or completion of a major project task, while the directed edges represent the tasks themselves. In a production project the initiation or completion of a production task (de-novo biosynthesis) are the events, while the edge that connects them is the production task. Each edge in the project graph is also equipped with a non-negative weight that equals the duration associated with the task. These durations are typically random. We tacitly assume that there are two unique nodes $s$, for start and $t$, for terminal, that mark the beginning and end of the project. The $s$ nodes instantaneously connect to all nodes that otherwise do not have entry nodes, and all the otherwise terminating nodes feed into the $t$ node. We call a path from start to end an $st$-path.

If a task cannot start before other tasks are completed, its start node is identified with the terminal node of these preceding tasks. The structure of the project graph thus captures the temporal precedence constraints among all production tasks and defines a partial order. Said differently, the nodes in the project graph function as ``AND'' gates, not allowing a new task to start (traverse an outgoing edge) before all predecessor tasks (preceding edges) are completed. This explains why the graph is both directed and acyclic, as it describes the progress in time of all the tasks in the project. 

Upon constructing the project graph and assigning durations to all its edges, the optimal \emph{project completion time} $T_c$ can be derived, by finding the duration of the longest $st$-path. This is because a successful completion of the project requires that all the activities are completed, and this only occurs when all the activities that belong to the longest $st$ path are completed. This path is known as the critical path; any delay in the activities belonging to it causes a global delay in the entire project. Other $st$ paths can have ``slack'' --- a duration gap between their completion time and $T_c$, allowing more flexibility in scheduling activities belonging to them.

Two archetypical examples of project graphs are a graph with a single dominant critical path (for example, a serial graph), and a balanced project graph, whereas the probability for a given $st$-path to be critical is as evenly distributed as possible, given the temporal constraints (for example, a fully parallel graph, for which the probability for any $st$-path to be critical is even). If fast completion time is a desirable goal, one can alleviate the constraints that causes a given critical path to be dominant. Iteratively repeating this constraint elimination process until convergence, results in a maximally balanced project \cite{TOC1,TOC2}, which is completed faster and is more accurate than the original project.


In Figure (\ref{fig2}) inset (C) we present how a project graph can be constructed for the self-replicating process described by Eq. (\ref{eq1}) and its corresponding catalytic reaction graph depicted in inset (A) of the same figure. The green circles indicate that these materials were produced by the previous replication round, which we denote by $k-1$. Thus, at initiation of replication, the three production processes marked by the three dotted, color filled, rectangles, can proceed concurrently. The corresponding doubling time $T_{\mu}$ will be the time when the slowest process among the three is completed: $T_{\mu}=max_{i \in \{1, \ldots, 3\}} T_i$.

We define the basic unit of self replication as the minimal amount of inventory (green circles) required for such parallel production. Inset (D) of the same figure demonstrates the effect of having exactly twice the inventory of inset (C). In this case, there are exactly two basic self replicating units, and the factory can proceed in parallel to produce four basic self-replicating units before splitting.

When the number of basic self-replicating units is below one, a delay is inevitable, because replenishing the missing inventory requires time. When the inventory level in not an integer, the integer part counts the number of parallel production processes. In the meantime, the non-integer part is actively synthesizing the missing materials. Thus, by correctly managing the substrate and catalyst inventory, parallel speedup can be obtained. If the factory is also balanced, such that the distribution of completion times for DNA replication (upper right dashed rectangle in Fig. (\ref{fig2}) inset (C)), catalysts and universal constructor replication (lower dashed rectangle in figure (\ref{fig2}) inset (C)) and material replication (upper left dashed rectangle in Fig. (\ref{fig2}), inset (C)) are approximately identical, and furthermore, the correlations between the different production activities is weak or null, then the distribution of the doubling time of the entire factory is well approximated by an extreme value distribution up to finite size corrections \cite{FiniteSizeCorrectionsToEVSPRL}. 

To achieve balance dynamically, cells often use end-product feedback inhibition. For example, in the biosynthesis of the amino-acid tryptophan, a metabolically costly amino-acid, excess tryptophan binds to the enzyme at the top-level of the dedicated metabolic pathway that produces it, thus down-regulating its own production \cite{Strayer}. A beautiful mechanism for translational control was discovered by Nomura \cite{Nomura2} where excess ribosome proteins that fail to bind to their targets, alternatively bind to their respective mRNA, thus stopping their own production. All these mechanisms use the excess end product, i.e. the part that was not consumed by downstream reactions, as a negative feedback signal that down-regulates its own production, resulting in a dynamically balanced production line.

\emph{The scheduling problem.--} In order for a biosynthetic task to be completed, the associated catalysts have to be allocated to the reaction along with the necessary input materials. Assuming the input materials are abundant, the following \emph{scheduling problem} arises --- how to assign tasks to processing units such that the completion time of the entire project is minimized? \cite{SchedulingBook}. In the noisy cellular milieu, we cannot expect scheduling algorithms to be precise. Instead, we ask what is expected from a randomized algorithm. Perhaps surprisingly, as long as the demand for catalysts is always met, the scheduling turns out to be optimal \cite{SchedulingBook}, as explained below.
To better characterize the scheduling problem it is useful to introduce the following parameters. The workload $T_w$ of a specific processor (catalyst) is defined as the sum over all durations of activities that require the processor. The workload represents the total demand for processors in terms of the processing time required. If there is only a single processor of a specific type, the optimal completion time is at least as great as its workload, since this processor has to perform all its tasks serially.
We define the average parallelism $p_c$ \cite{avgpc} for a particular processing unit by dividing its associated workload $T_w$, by the critical path duration $T_c$, and rounding up:
\begin{equation}\label{AverageParallelism}
p_c=\lceil \frac{T_w}{T_c} \rceil.
\end{equation}
The average parallelism is an estimate for the minimum number of processors required to allow for optimal completion time. Another interpretation of the average parallelism comes from considering a completely parallel graph, which is a graph composed of many non-overlapping $st$-paths. In this graph, $p_c$ is simply the number of effective parallel independent $st$-paths. When two $st$ paths are connected by an activity edge, their completion time is correlated, since activities downstream of the edge in one path cannot be completed before upstream activities in the other path are completed. Due to these correlations, calculating the distribution of project completion times can be hard.

There are two different ways in which a mismatch in the number of catalysts adversely affects the completion time. The first is lack of catalysts with respect to the demand, and the second is excessive abundance. If catalysts are missing, it means that there are bio-synthesis tasks that idle, until catalysts become available to perform them. If these delayed tasks are not residing on the critical path, this idling may not affect the overall completion time, as long as the delay is shorter than the ``slack'' which is defined as the minimum duration gap between the critical path and all the $st$-paths that share these idling edges. If on the other hand, the idling edges are on the critical path, the slack is zero, and the delay immediately adds up to the completion time of the entire project.

When, on the other hand, there is an excess of catalysts, then whenever there is a demand for them, it can readily be met, provided that the pool of catalysts is obeying the simple premise that catalysts never idle if there is an available task (this circumstance is known as greedy scheduling). Thus, an excess of catalysts with respect to demand is a good thing, as it allows for optimal completion time, determined only by the critical path duration. Moreover, this excess also allows for proper balancing in the presence of random delays that affect the actual demand. However, at some point, having an excess of catalysts can become sub-optimal, since without further control, a greedy scheduling policy can deplete either the input resources or the catalytic pool, by allocating catalysts to reactions that have randomly completed earlier, causing a speedup in the progress of certain activities at the expense of others. Finally, we note that the amount of excess is also restricted  because of cost and space considerations.

The duration of the activities in a project are stochastic variables. Hence the project workload $T_w$, the critical path duration $T_c$ and the average parallelism $p_c$, are also stochastic. We can thus generalize $p_c$ to a new parameter, $\tilde{p}_c$, which is the minimal number of processing units required for optimal completion with a given confidence level. For example, if the distribution of the average parallelism is exponential, a $95\%$ confidence level will require $\tilde{p}_c=3 \bar{p}_c$, where  $\bar{p}_c$ is the average $p_c$ over many random realizations of durations (edge weights) for a given project graph. Thus, when the number of processors exceeds $\tilde{p}_c$ then in $95\%$ of the cases, the project will be completed at the optimal time $T_c$ \cite{SchedulingBook}.

\begin{figure}
\centering
\includegraphics[width=0.8\columnwidth]{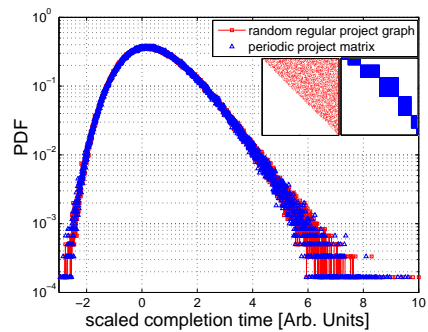}\\
\caption{(Color online) Probability distribution function of the scaled project completion time for two types of project graphs with $N=512$ and $N=1204$ nodes respectively. Blue triangles --- a graph with a periodic structure. Solid red line with red squares --- a random regular sparse project graph.}\label{fig3}
\end{figure}

\emph{basic unit of self-replication.--} Upon introducing the average parallelism parameter, it is natural to classify self-replication in relation to the level of parallelism. To facilitate this definition we first define the processor demand vector $\vec{d}^c$, such that
\begin{equation}\label{AverageParallelism2}
d^c_j=\sum_{i=1}^{n_R} C_{ij}.
\end{equation}
Similarly, we can define the demand for consumable material inputs (substrates) $\vec{d}^s$:
\begin{equation}\label{AverageParallelism3}
d^s_j=\sum_{i=1}^{n_R} S_{ij},
\end{equation}
where the catalyst input matrix, $C$, of size $n_R \times n_C$, has an entry $(C)_{ij}$ equals to the number of catalysts of type $j$ required by reaction $i$ ($n_R$ is the total number of reactions in the reaction graph, and $n_C$ is the number of catalyst types). The matrix $S$ is defined similarly i.e. $S_{ij}$ is the number of substrates of type $j$ consumed by reaction $i$. The total demand vector is then $\vec{d}=\vec{d}^c+\vec{d}^s$. Let $\vec{n}(t=0)$ be the occupation number vector whose elements account for the number of processing units per type at time $t=0$. If any of the elements in $\vec{n}(0)$ is less than the corresponding element in $\vec{d}$ then the process is not fully parallel, since some processors are missing. In order to achieve parallelism, and in order to balance production, the missing processors should be replenished. When, on the other hand, $\vec{n}(0)=p \times \vec{d}$, with $p=1$ the different de-novo synthesis processes are decoupled for the first time, which means that there are enough processing units to allow them to work in parallel. It follows that the doubling time $T_{\mu}$ defined as $\vec{n}(T_{\mu}) = 2\vec{n}(0)$ is given by the maximum among all the different production tasks:
\begin{equation}\label{AverageParallelism22}
T_c=\max_{i \in \{1, \ldots ,n_R\}} T_i.
\end{equation}
In general, the durations $T_i$, $i \in \{1, \ldots ,n_R\}$ are not independent as explained below.

If the initial state is an integer multiple $p>1$ of the demand vector, and the factory is balanced, the resulting dynamics is of exponential (and balanced) growth, with $p$ parallel self-replicating basic units. In the absence of correlations, the resulting distribution of completion time belongs to the Extreme Value Statistice (EVS) limit distribution. In practice, in order to keep the progress well-balanced, some form of coupling still exists between the basic self-replicating units. For example, in \textit{E. coli}, the production of ribosomal proteins and ribosomal RNA is co-regulated. DNA replication is also regulated to match the overall growth rate.

In the serial limit, the number of basic self-replicating units present is less than one. A lag phase during which the missing processors are synthesized is inevitable. Incidently, we note that the question of which processors to create first is also of interest, as order can have a significant impact on both the rate of recovery as well as the metabolic cost. For example, germinating spores or starved cells recently moved to a rich medium, both have to replenish and recharge their metabolic and catalytic pools and thus can be considered to be transiently in the serial phase.

\begin{figure}
\centering
\includegraphics[width=1 \columnwidth]{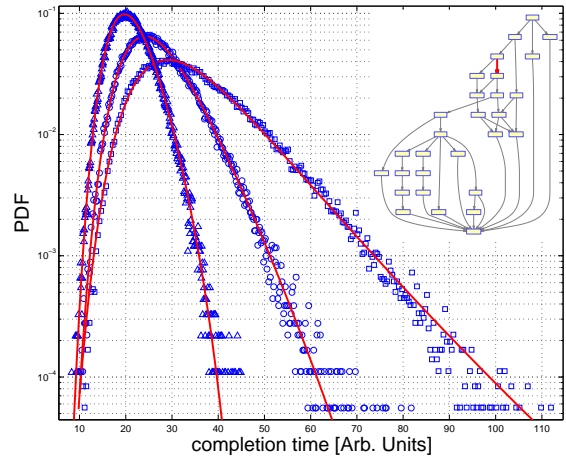}\\
\caption{(Color online) Balanced vs. unbalanced projects. The project graph, depicted in inset (a) is adapted from the Nomura assembly map for the $30S$ ribosome subunit \cite{Williamson,Nomura1}. We performed a Monte-Carlo (MC) simulation with $10^6$ iterations. In each iteration, we assigned random durations for each edge, drawn from a $\Gamma$ distribution with a shape parameter $2$ and a scale parameter, $1$. For each random realization we calculated the optimal completion time $T_c$, which is the first time the 30S subunit can mature. The probability density function (PDF) calculated based on the MC run is presented in blue triangles. The blue squares and circles correspond to the case where a statistical delay by a factor of $2$ (circles) and $4$ (squares) in the binding of ribosomal protein S4 to its rRNA target site, relative to its baseline value (triangles). This delay causes an imbalance in the graph, deferring all subsequent activities. Due to the periodic-like graph topology \cite{remarkQuasi} the delay did not alter the functional form of the distribution, but rather changed its three defining parameters. We used a maximum likelihood estimator (MLE) to find the parameters of the generalized extreme value (GEV) distribution for each Monte-Carlo generated completion time PDF, which is a good approximation to the data at this range. The resulting MLE fits are depicted by solid red lines. Inset (a) shows the project graph. The red edge is the edge which we delayed by a factor of $2$ or $4$ relative to its initial value (binding of ribosomal protein $S4$ to the rRNA backbone).} 
\label{fig1b}
\end{figure}


\emph{Calculating the distribution of the longest path for balanced growth.--} Consider a given project graph, and assume that the duration of each task is a random variable distributed according to some distribution which is at least exponentially decaying at large times. To calculate the project completion time we introduce the exponentially weighted adjacency matrix of the project graph, which is defined as
\begin{equation}\label{ExpAdjMatrix}
  M_{ij}=e^{\beta T_{ij}},
\end{equation}
where $T_{ij}=T(\{ e \}$ is the duration of a task $e=(i,j) \in E$, $T_{ij}=-\infty$ if $(i,j)$ is not an edge in the graph and $\beta \gg 1$ is a large positive number. We arrange the indices of the nodes such that the first node $(i=1)$ represents the project start, and the last node $i=n$ represents the project completion. We also introduce an auxiliary edge, $M_{n1}=1$ which can be considered as a resetting edge that returns the project back to start, upon the completion of the last task, thus mimicking the division event. We denote the project matrix with the auxiliary node by $M_c$, where the dependence on $\beta$ is suppressed for brevity. The matrix trace $Tr(M_c^k)$ is composed of a sum over all loops with $k$ activities, exponentially weighted by the total duration of each k-loop. In the limit $\beta \rightarrow \infty$ , the leading contribution is the longest path with $k$ steps. We further divide the trace by $k$, since there are $k$ equally contributing nodes on each loop of length $k$, depending on where we locate the loop's start. Summing over all $k$'s we obtain a discrete path integral which is a function of $\beta$. Dividing by $\beta$ and taking the limit $\beta \rightarrow \infty$ we obtain
\begin{equation}\label{Tc}
  T_c=\lim_{\beta \rightarrow \infty} \frac{1}{\beta}\sum_k \frac{Tr(M_c^k)}{k}=\frac{1}{\beta}\sum_k \sum_l \frac{\lambda_l^k}{k},
\end{equation}
since $Tr(M_c^k)=\sum_{l=1}^{n} \lambda_l^k$, where $\lambda_l$ is the $l^{th}$ eigenvalue of $M_c$. To evaluate this sum, we add a regularizing parameter $u$ and perform the sum over $k$:
\begin{equation}\label{Tc22}
  T_c=-\lim_{\beta \rightarrow \infty} \frac{1}{\beta} \sum_l \log (1-u\lambda_l) = -\lim_{\beta \rightarrow \infty} \frac{1}{\beta}\log |I-uM_c|,
\end{equation}
where $|I-uM_c|$ is the determinant of $I-uM_c$. $T_c$ is then realized as the regularized $u=1$ value. 
Equation (\ref{Tc22}) also shows that the project completion time is an asymptotic spectral property of the project graph adjacency matrix. To obtain the distribution of completion times, one has to calculate the probability distribution of $|I-uM_c|$ with a specified ensemble of random project matrices. The randomness can be for a fixed graph topology with random weights (durations), or can include the graph topology. In the next sections, we focus on the former, and solve for the distribution of completion times for a class of project matrices that describe a balanced production line.

\emph{Balanced self-replication has a periodic project matrix.--} Recall that the vertex set $V$ of a directed periodic project graph has a unique property that it can be partitioned into $q$ disjoint sets, $V=V_1 \cup V_2 \cup \ldots \cup V_q$, such that every edge that originated in $V_i$ ends up in $V_{i+1}$ with $V_{q+1}=V_1$ \cite{digraphBook}.

In a balanced self-replicating factory, the progress in production of each processor type is controlled such that the production front --- the plane that best fit the occupation number vector $\vec{n}(t)$, progresses uniformly (i.e. parallel to itself), up to some statistical tolerance $\delta \vec{n}$. Hence, starting from the occupation number vector $\vec{n} \pm \delta \vec{n}$ the occupation number vector grows statistically uniformly to $2 \vec{n} \pm 2\delta{\vec{n}}$ and back to $\vec{n} \pm \delta \vec{n}$ upon division. Alternatively, we can think of balanced production as being composed of stages. Activities in the next stage will not start before certain activities in the previous stage are completed. Both descriptions lead to a periodic project graph, provided we close the terminal and start nodes with a ``reset'' edge that plays the role of division.

The project completion time of an irreducible periodic matrix project can be calculated analytically, due to the discrete rotational symmetry of the adjacency matrix spectrum. The eigenvalues satisfy $\lambda_l=|R_l| \mathbf{e}^{\frac{2 \pi i l}{q}}$. The maximal eigenvalue is real, and has a value which we denote by  $\lambda_{max}=R_{max}$. Plugging in Eq. (\ref{Tc22}) we obtain
\begin{eqnarray}\label{TcPeriodicMatrix}
 T_c=-\lim_{\beta \rightarrow \infty} \frac{1}{\beta} \sum_l \log (1-u|R_l|) = \nonumber \\
 -\lim_{\beta \rightarrow \infty} \frac{1}{\beta}\log R_{max},
\end{eqnarray}
The distribution of the maximal eigenvalue of a random Wishart matrix was recently shown to belong to the Frechet class when the entries are power law distributed \cite{BenArous}. As shown in the SI, this implies that the maximal eigenvalue of $M_c^t M_c$ is distributed according to the Frechet distribution. Thus, the desired distribution of $T_c$ is proportional to $\log R_{max}$ which follows the log-Frechet distribution. In practice, we find that the distribution is well approximated by the generalized Gumbel distribution \cite{BHP} up to $6$ orders of magnitude in the probability axis. Note however that an analytical formula for the log-Frechet distribution is derived in the SI. We also present numerical evidence that the realm of validity of the predicted universal form of the optimal project completion time distribution extends beyond periodic project graphs, suggesting a possible new type of universality class for extremely non-symmetric random matrices.

\begin{figure}
\centering
\includegraphics[width=1 \columnwidth,height=1 \columnwidth]{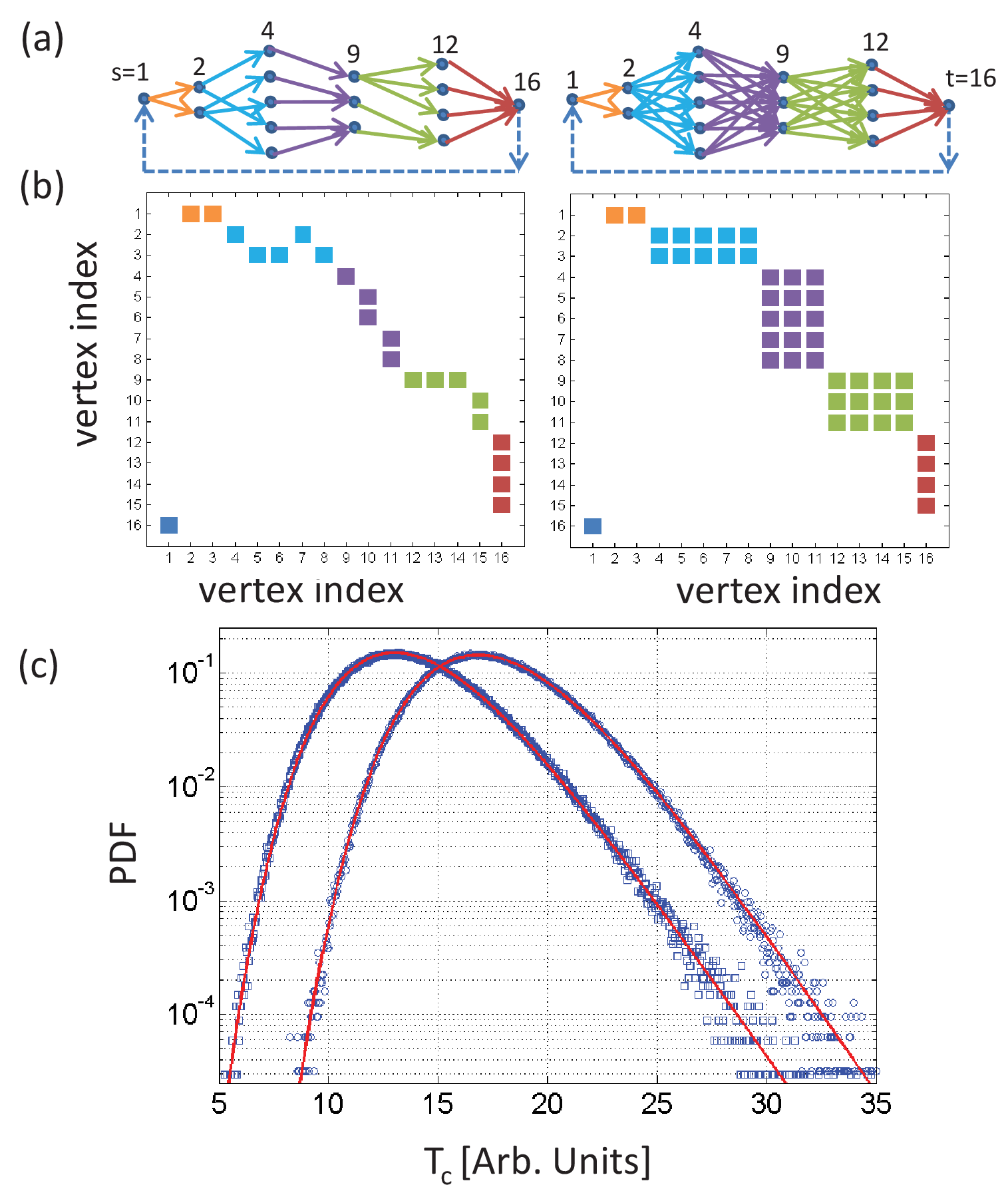}\\
\caption{(Color online) Production in well defined stages leads to balanced growth. This figure illustrates this point by presenting two balanced project graphs with five stages (represented by five different edge colors). In the right-hand side we present a tightly controlled project graph, where the edges that connect each stage form a complete bipartite graph. This is the tightest control possible; no stage can start prior to the completion of the previous stage. In the left-hand side, we present a less stringent project obtained by pruning the edges of the right-hand side graph. Inset (a) The project graph. Note the ``division'' task (dashed backward arrow) that resets the factory back to the start node ($s=1$) upon completion of the last stage. (b) Adjacency matrices $A$ of the two project graphs. Non-zero entries are colored according to the color of the edge they represent. For example, the entry $A_{12,16}$ that represents the edge from vertex $12$ to the terminal node $t$ ($t=16$) is marked by a red square. In inset (c) the distribution of project completion times obtained from a MC simulation with $10^6$ random realizations of task durations (edge weights) of the two graphs is presented. The blue squares are the results for the left-hand side graph, and blue circles are for the right-hand side graph. The red lines are the fitted universal (log-Frechet) theoretical curve predicted by our calculation.} 
\label{BalProd}
\end{figure}

In Fig. (\ref{BalProd}) two example for balanced project graphs with a varying degree of control over project progress are presented. The project graph, its incidence matrix, the Monte-Carlo simulated distribution of optimal completion times, and the theoretically predicted universal shape are presented in insets (a)-(c). 

\begin{figure}
\centering
\includegraphics[width=1 \columnwidth]{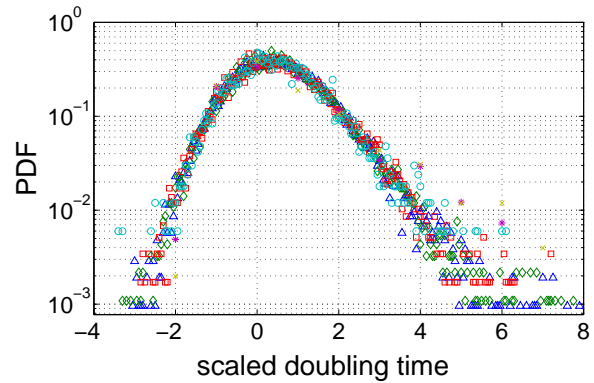}\\
\caption{(Color online) Data collapse of the measured doubling times on the log-Frechet probability distribution function. Data from \cite{jun} are \textit{E. coli} cells growing in LB media at $37$ $^{\circ}C$, at an average doubling time of $21$ minutes. Although differences between the original distributions existed in their respective variance and skew, the shape after the collapse is universal and agrees with the predictions of optimal completion time when growth is balanced (or quickly balances). Additional data from \cite{Elf} where \textit{E. coli} was grown on different media (M9 minimal media with or without supplemented amino-acids) also collapse, but less accurately, perhaps due to the significantly smaller sample size.}\label{JunElf}
\label{datc}\end{figure}

To test our prediction of optimal doubling time we analyzed three data sets of \textit{E. coli} growing in different media. The first data-set obtained from \cite{jun}, has an unprecedented number of division measurements, out of which we filter roughly $80000$ clean divisions of either mother, daughter or granddaughter \textit{E. coli} cells growing rapidly in Luria-Bertani (LB) broth at $37$ $^{\circ}C$ within a narrow micro-fluidic canal, at an average doubling time of $21$ minutes \cite{jun}. The doubling time $T_{\mu}$ was calculated by the formula
\begin{equation}
T_{\mu}=\frac{T_{div}}{\log_2 R(T_{div})},
\label{tmu}\end{equation}
where $T_{div}$ is the inter-division time, and $R(T_{div})$ is the length ratio (ratio of the length of the cell at division divided by its initial length). We compared two methods to calculate $T_{\mu}$. The first is described by Eq. (\ref{tmu}) and the second is by fitting an exponent to the measured growth curve. We obtained practically identical doubling times in the two methods. After filtering noisy data, we calculated the observed doubling time distribution and used standard least square fitting to find the location and scale parameters of the log-Frechet distribution. We then scaled and centered the doubling time and plotted the PDF as a function of this scaled variable. A clear data collapse of all the different types of cells on the log-Frechet distribution is evident in Fig. \ref{JunElf}. A second data set with $1500$ division events from ref \cite{Elf}, of \textit{E. coli} growing on minimal (M9) medium with or without additional amino-acids was analyzed. The data also collapse to the log-Frechet curve, yet this collapse is more noisy, partially because of a finite size effect.

\emph{Conclusion and outlook} We studied the distribution of project completion time of a self-replicating factory. Invoking PERT and the notion of balanced growth \cite{CooperBook} a periodic structure for the project graph of a self-replication production process was suggested. The distribution of optimal completion times for this graph structure was calculated and found to have a universal shape --- the log-Frechet distribution. Contrary to other combinatorial optimization problems, finding the optimal scheduling in a self-replicating factory can be trivialized by requiring that the inventory of processing units is in excess relative to the demand. In that case, the distribution of optimal project completion times is also the distribution of actual completion times. In order to maintain such a stoichiometric balance requires some level of control over the rate of production of different types of processing units, leading to a global structure of a periodic project graph. To corroborate our model, we analyzed \textit{E. coli} growth data from LB media \cite{jun} and from M9 with and without supplemented amino-acids \cite{Elf}, and showed that the data collapses on our predicted single universal shape using two fitting parameters that are linearly related to the mean and variance of the actual doubling time.

In the future, it would be interesting to experimentally test our hypothesis by measuring individual cell's replicative buffer size and the inter-division time. Starting a new round of DNA replication is in one-to-one correspondence with initiating more ribosome production lines i.e. rRNA operons. Thus, the replicative buffer should correlate with the number of origins of replication in a wild-type cell in normal conditions. Furthermore, by using mutant strains with e.g. a defects in the coordination of DNA replication with the growth rate, or with several  deleted ribosomal operons, a deviation of the doubling time distribution from its universal form is expected. In contrast, the distribution of doubling times for wild type \textit{E. coli} is expected to collapse on the same curve as in Fig. (\ref{datc}), irrespective of the type of environment, as long as all transients have settled and the environment is held fixed.

I would like to thank Noya Shilo, Yoav Sagi, Oren Raz, Or Meir, Shachar Lovett, Shlomi Kotler, Danny Ben-Zvi, BingKan Xue, Ori Parzanchevsky, Alon Wagner, Tsvi Tlusty, Benjamin Bratton, Edinah Gnang, Avi Wigderson, Arnold Levine, Hanna Salman, Alber Libchaber and John Hopfield for helpful comments and illuminating discussions. I would like to thank Suzanne Christen for her comments and suggestions. Special thanks to Itzhak and Rachel Hen and Lital and Amit Sever for their support. The support of the Eric and Wendy Schmidt membership in biology and the Janssen fellowship are gratefully acknowledged.


\begin{thebibliography}{22}
\bibitem[{rem()}]{remarkQuasi}
\bibinfo{note}{We looslely define a periodic-like graph as a graph whose
  spectrum in the complex plane is close to an periodic graph spectrum. Thus,
  there exsit a small parameter $\epsilon$ much smaller than the gap in the
  spectrum, such that all eigenvalues of the periodic-like matrix are
  $\epsilon$ away from a the eigenvalues of a periodic matrix of equal size.}
\end{thebibliography}

\end{document}